\documentclass[journal=langd5,manuscript=article]{achemso}

\usepackage{mathptmx}
\usepackage[version=3]{mhchem} % Formula subscripts using \ce{}
\usepackage[T1]{fontenc}       % Use modern font encodings

\author{Hongjie An}
\email{hjan@ntu.edu.sg}
\affiliation[Nanyang Technological University]{Cavitation Lab,
  Division of Physics and Applied Physics, School of Physical and
  Mathematical Sciences, Nanyang Technological University, Singapore}
\author{Beng Hau Tan}
\affiliation[Nanyang Technological University]{Cavitation Lab,
  Division of Physics and Applied Physics, School of Physical and
  Mathematical Sciences, Nanyang Technological University, Singapore}
\author{Claus-Dieter Ohl}
\email{cdohl@ntu.edu.sg}
\affiliation[Nanyang Technological University]{Cavitation Lab,
  Division of Physics and Applied Physics, School of Physical and
  Mathematical Sciences, Nanyang Technological University, Singapore}

\title[An \textsf{achemso} demo]
  {Distinguishing nanobubbles from nanodroplets with AFM: the influence of vertical and lateral imaging forces}

\abbreviations{HOPG}

\begin{document}

\begin{tocentry}

% \begin{figure}
\includegraphics[width=0.75\columnwidth]{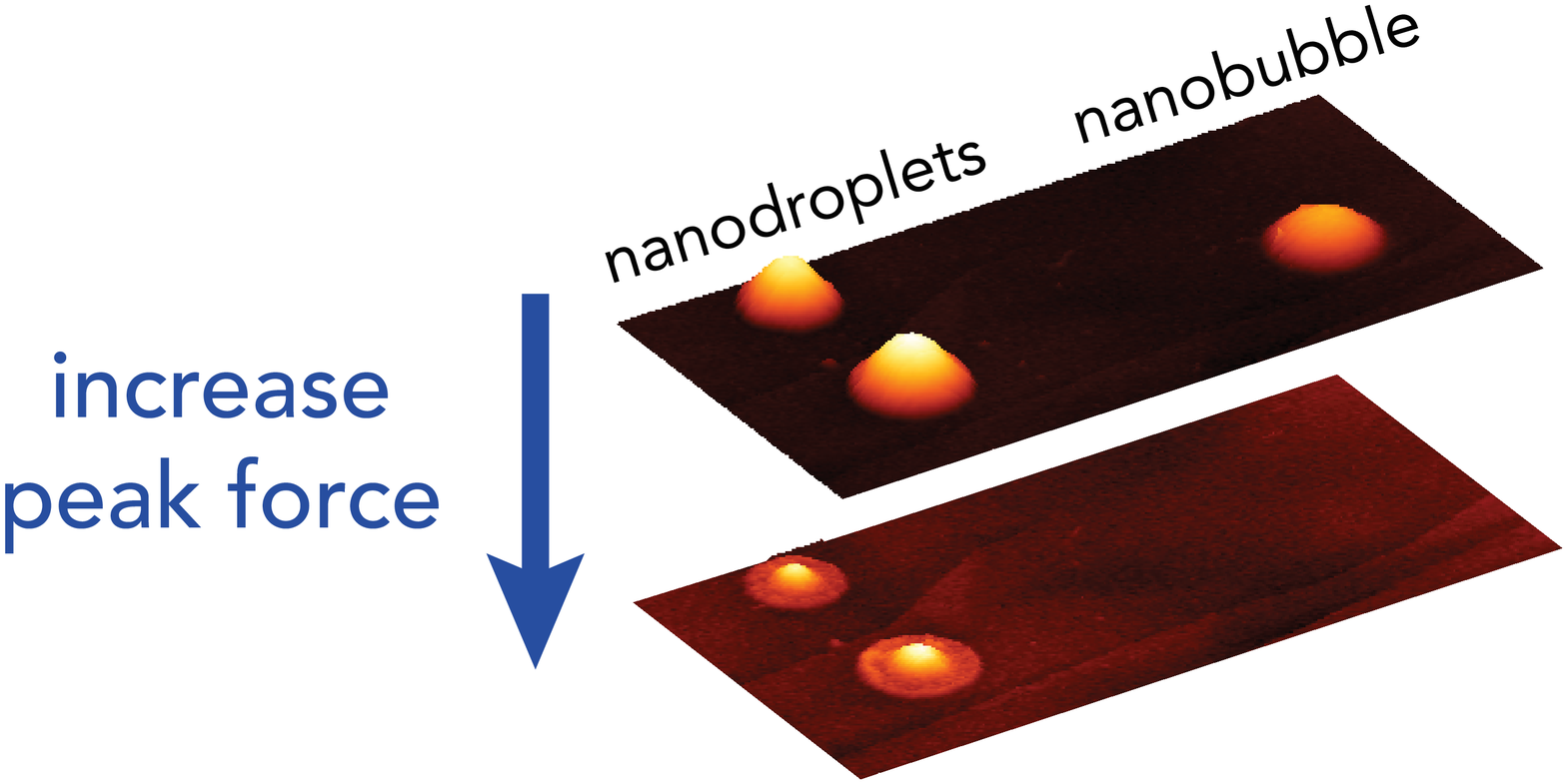}
% \end{figure}

% Some journals require a graphical entry for the Table of Contents.
% This should be laid out ``print ready'' so that the sizing of the

% Inside the \texttt{tocentry} environment, the font used is Helvetica
% 8\,pt, as required by \emph{Journal of the American Chemical

% The surrounding frame is 9\,cm by 3.5\,cm, which is the maximum
% permitted for  \emph{Journal of the American Chemical Society}
% graphical table of content entries. The box will not resize if the
% content is too big: instead it will overflow the edge of the box.

% This box and the associated title will always be printed on a
% separate page at the end of the document.

\end{tocentry}

\begin{abstract}
  The widespread application of surface-attached nanobubbles and
  nanodroplets in biomedical engineering and nanotechnology is limited
  by numerous experimental challenges, in particular, the possibility
  of contamination in nucleation experiments. These challenges are
  complicated by recent reports that it can be difficult to
  distinguish between nanoscale drops and bubbles. Here we identify
  clear differences in the mechanical responses of nanobubbles and
  nanodroplets under various modes of AFM imaging which subject the
  objects to predominantly vertical or lateral forces. This allows to
  distinguish between nanodroplets, nanobubbles, and oil covered
  nanobubbles in water.
\end{abstract}

\section{Introduction}
The nucleation of micro and nanoscale liquid drops and gaseous bubbles
on surfaces has gathered significant interest in recent years for
numerous applications \cite{craig_very_2010, seddon_nanobubbles_2011,
  lohse_surface_2015}. Nanodroplets are useful for biomolecular
analysis and microfluidic reactors \cite{zhang_formation_2015}, while
the decoration of surfaces with bubbles significantly reduces fluid
drag in microchannels \cite{karatay_control_2013}. Both nanobubbles
and nanodroplets are produced by exchanging water with an organic
solvent, such as ethanol, over a suitable surface
\cite{craig_very_2010, seddon_nanobubbles_2011, lohse_surface_2015,
  zhang_formation_2007}. The nucleation mechanism of nanobubbles and
nanodroplets appears to depend on the substantial difference in the
solubility of gas (nanobubbles \cite{german_interfacial_2014,
  chan_tirf_2012}) or the desired liquid (nanodroplets
\cite{zhang_formation_2007}) in the water and solvent.

Given the similarity in their nucleation recipes, the contamination of
exchange liquids may lead to the nucleation of both nanobubbles and
nanodrops within a single experiment. It was recently reported that
using disposable medical plastic syringes and cannulas to deliver
liquids lead may lead to the contamination of nucleation experiments
\cite{berkelaar_exposing_2014, an_wetting_2014,
  carr_plastic-syringe_2012, buettner1991free}. This contamination
arises from the use of polydimethylsiloxane (PDMS) or other
biologically-inert silicone oils to lubricate syringes and needles, in
order to reduce discomfort during topical injections
\cite{siniawski_method_2015, curtis_chapter_2013}.

The issue of contamination contributes to reproducibility issues and
conflicting results in the field of nanobubbles. It has been noted
that the height and size distributions of nanobubbles characterised
with atomic force microscopy (AFM) varies drastically between research
groups \cite{ seddon_nanobubbles_2011, lohse_surface_2015}, even
though only widely-available liquids and atomically smooth substrates
like highly oriented pyrolytic graphite (HOPG) are used in most such
experiments.

Given the ease of contamination, experimental methods to distinguish
polymeric liquid nanodroplets and gaseous nanobubbles are of urgent
interest \cite{ seddon_nanobubbles_2011, lohse_surface_2015}. Recently
significant progress in this area has been made. Chan {\em et al.}
\cite{chan_collapse_2015} used fluorescence microscopy to show that
moving a contact line over a nanobubble deflated it upon exposure to
the ambient atmosphere, while the contact line pinned strongly on a
nanodroplet. Seo {\em et al.}~\cite{seo_phase_2015} distinguished
between dye-covered bubbles and drops by identifying differences in
fluorescence signals. However, the two tests either permanently alter
the interfacial chemistry of the objects or destroy them upon
identification. Moreover, the fact is that the vast majority of the
literature characterise nanobubbles exclusively with AFM. It is
therefore important to be able to make the distinction between
nanobubbles, nanodroplets and contamination using AFM in particular,
so as to allow researchers in this field an opportunity to verify that
their previously-published nucleation protocols are not contaminated. 

In this article, we demonstrate that surface nanobubbles and
nanodroplets on an atomically flat HOPG substrate can be
differentiated through careful, non-destructive AFM
characterization. We find distinct differences in the response of
nanobubbles and nanodroplets in (a) PeakForce mode, in which the
vertical imaging force can be carefully controlled; (b) contact mode,
which delivers an invasive lateral force on the objects; and (c) force
spectroscopy.

\section{Results and discussion}

\subsection{Generation of nanobubbles and nanodroplets}
\begin{figure}
\includegraphics[width=\columnwidth]{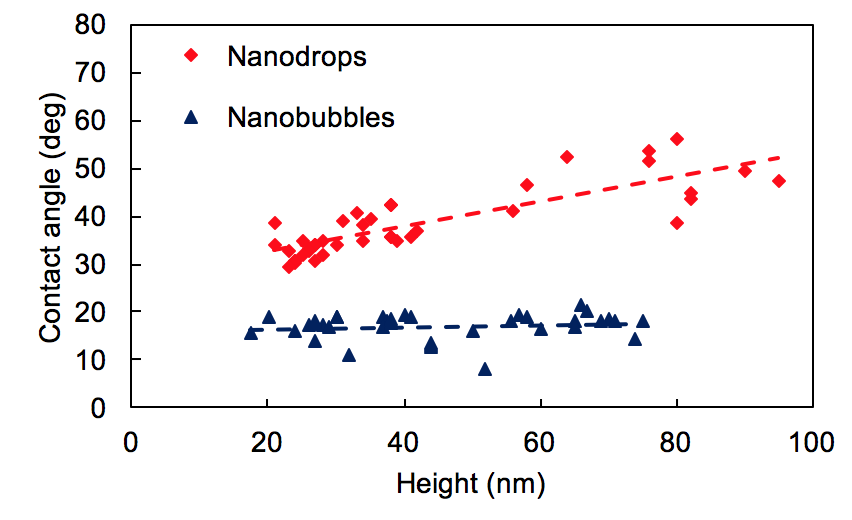}
\caption{\label{fig1} Nanoscopic contact angles of nanobubbles
  ($n=36$) and nanodroplets ($n=40$) as function of height. The angles
  are calculated by a least-square fit to a spherical cap. The
  nanoscopic contact angles of nanodroplets (red diamonds) increases
  with height, while nanobubbles (blue triangles) have a contact angle
  of $\approx 20^\circ$ which is independent of height. }
\end{figure}

To eliminate external contamination, we used unambiguous methods to
nucleate nanobubbles and nanodroplets, avoiding solvent exchange or
plastic syringes. Hydrogen nanobubbles were electrolytically generated
on HOPG, while nanodroplets were introduced by directly depositing 
 a dilute solution of PDMS (see Materials and Methods).

 A baseline distribution of sizes and contact angles was established
 by imaging nanobubbles and nanodroplets separately with tapping mode
 AFM. The nanoscopic contact angles of the objects were calculated by
 least-square fits of the cross-sectional profile to a spherical
 cap. As shown in Figure 1, the nanobubbles' fitted contact angles
 were virtually height-independent ($\theta \approx 20^\circ$),
 whereas the nanodroplets' contact angles increased
 ($30^\circ<\theta<60^\circ$) with height. The contact angles are
 measured from the gas phase for nanobubbles and from the oil phase
 for the nanodroplets.

\subsection{Morphology of nanobubbles and nanodroplets under varying
  peak force}

\begin{figure}
\includegraphics[width=\columnwidth]{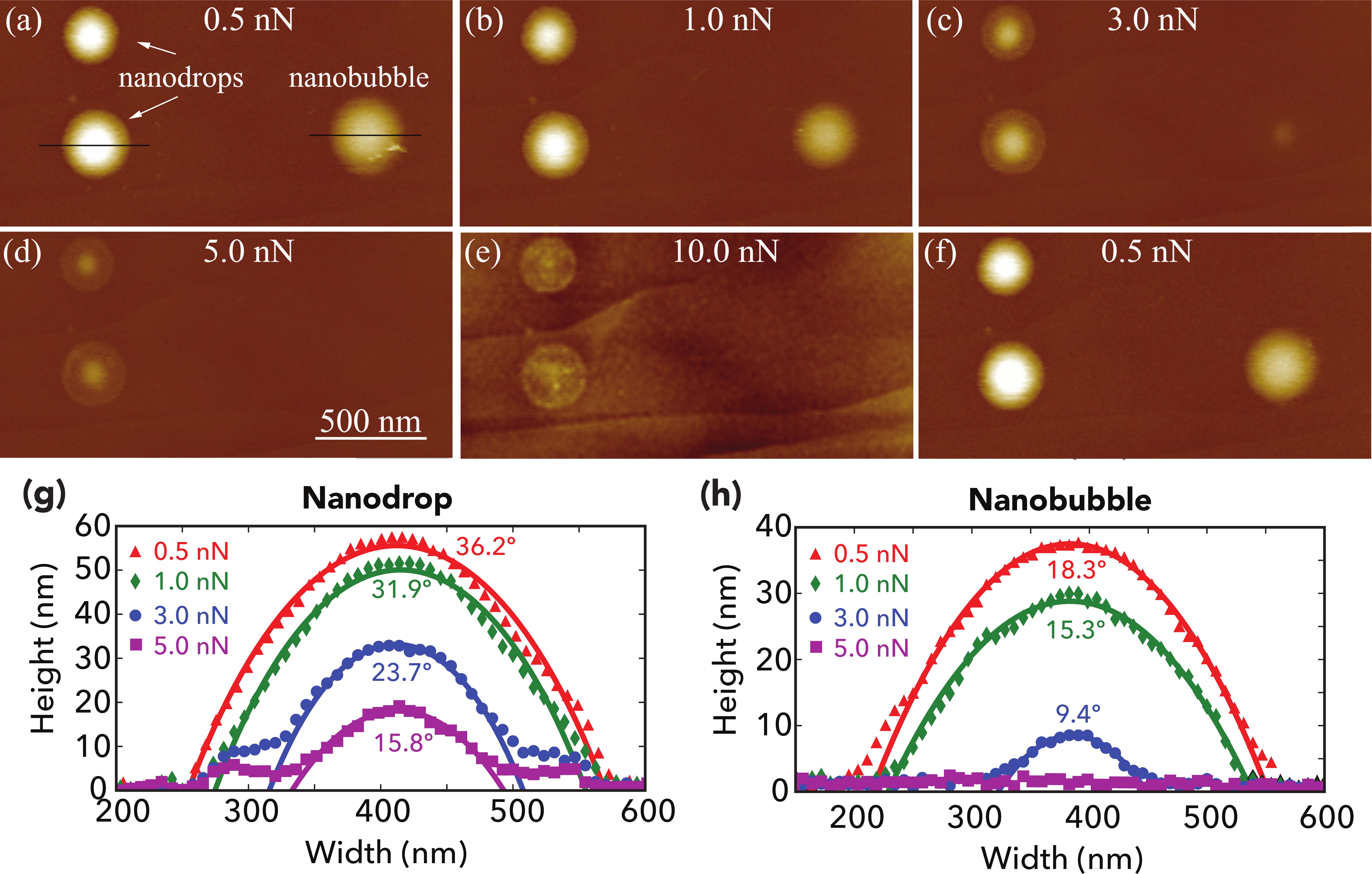} %1.7
\caption{\label{fig2} AFM height images of PDMS nanodroplets and a
  nanobubble in PeakForce mode. (A)-(E) Successive AFM images captured
  for peak forces $F_p = $ 0.5, 1.0, 3.0, 5.0 and 10.0 nN. A final
  scan was taken at $F_p = 0.5$ nN, showing that the objects were not
  destroyed by the scanning. Scan size: 2 $\mu$m $\times$ 1
  $\mu$m. Height scale: 50 nm for (A) - (D) and (F); 5 nm for (E).
  (G) Cross-sectional profiles of the bottom-left nanodroplet in
  (A)-(F). The nanodroplet adopts a sombrero shape at $F_p > 3$ nN, with
  a molecular layer of $\sim$ 1 nm. (H) Cross-sectional profiles of
  the nanobubble. In (G) and (H) contact angles are indicated, and the
  lines are least-square fits to a spherical cap.}
\end{figure}

The most direct way to compare nanobubbles and nanodroplets is to
introduce them onto a common substrate. We first electrolysed water to
generate bubbles, and later introduced oil drops by adding dilute oil
solution, allowing both types of objects to appear on the same scan
(Figure 2). We selected an area containing two nanodroplets and one
nanobubble with similar lateral radii. To observe the effects of
varying the imaging force on drops and bubbles, we used PeakForce
tapping mode AFM. In this mode, the maximum vertical force on the
cantilever is used as a feedback mechanism, allowing soft objects to
be imaged with a {\em peak force} $F_p$ that is typically of the order
0.1 to 1 nN, but can be controlled to an accuracy of $\sim$10 pN
\cite{yang_imaging_2013, walczyk2013effect, zhao_mechanical_2013}.

We then performed successive AFM scans at $F_p = $ 0.5 to 10 nN
(Figures 2A-E). The morphology of nanodroplets and nanobubbles changed
significantly with increasing $F_p$. At 0.5 nN, the height profiles of
nanodroplets (Figures 2A and G) and nanobubbles (Figures 2A and H)
were well-fitted to spherical caps. With increasing $F_p$, the
nanobubble in Figure 2a became smaller in base radius and height,
before disappearing completely at $F_p = 5.0$ nN. A nanodroplet also
shrank in base radius and height with an increase in $F_p$, but at
$F_p > 3$ nN appeared as a {\em sombrero}: a spherical cap sitting on
a flat molecular layer. Remarkably, the molecular layer was highly
resistant to vertical loading and remained in the height image at up
to $F_p = 20 $ nN, maintaining a thickness of 1-2 nm. Molecular layers
of $\sim$1-2 nm thickness at the contact lines of spreading PDMS drops
have previously been observed by ellipsometry
\cite{heslot_molecular_1989}, though this layering is still not well
understood. Finally, we note that neither bubbles nor drops were
physically moved or destroyed by PeakForce imaging. When $F_p$ was
reduced from 10.0 to 0.5 nN, all the objects restored their original
heights, base radii, and positions (Figure 2F).

The way we created the objects on the system -- with bubbles created
first, and drops introduced later -- leaves open the possibility that
the bubbles may have been coated with oils. To investigate this
scenario, AFM scans were taken in PeakForce mode before and five hours
after dilute oil solution was introduced into the system (Figure S1 in
the Supporting Information). Some bubbles exhibited an increase of
contact angle from 18-20$^\circ$ to 30-54$^\circ$ (Table S1, Figures
S1-2). As surface nanobubbles are stable against diffusion for several
weeks without significant changes to morphology, the difference in
contact angle must arise only from the change in interfacial energy
balance at the nanobubble's three-phase line due to oil depositing on
top of the nanobubbles. When $F_p$ was increased
incrementally, nanobubbles no longer disappeared from the scan image
at $F_p=$ 5.0 nN but maintained a nanometric layer (see Figures S1-2),
similar to the oil nanodroplets.

The presence of a nanometric layer under strong AFM scanning is a
tell-tale sign of contamination, resolving a contradiction raised by
three recent papers imaging nanobubbles in PeakForce mode. Zhao {\em
  et al.}  \cite{zhao_mechanical_2013} and Yang {\em et al.}
\cite{yang_imaging_2013} found that nanobubbles disappeared from the
scan image at $F_p =$ 1.25-2 nN, restoring their original heights when
$F_p$ was reduced to 0.2 nN. This agrees with our results. On the
other hand, Walczyk {\em et al.}  \cite{walczyk2013effect} suggested
that nanobubbles ($\sim$20 nm height, $\sim$100 nm width) remained
visible when $F_p$ reached 27 nN. While the vast majority of work in
the literature nucleated nanobubbles by solvent exchange or
electrolysis \cite{craig_very_2010, seddon_nanobubbles_2011,
  lohse_surface_2015}, they were able to nucleate the objects simply
by delivering water with a disposable syringe. Our results suggest
that the robust layers visible at high peak forces in Walczyk
{\em et al.} \cite{walczyk2013effect} were not nanobubbles, but
polymeric contamination originating from plastic syringes.

\subsection{Invasive contact mode scanning}

Soft objects are usually imaged in non-invasive tapping or PeakForce
mode AFM, rather than the conventional contact mode. In these modes,
the tip height is adjusted rapidly (once in 100 $\mu$s in PeakForce
mode \cite{kaemmer2011}) by a feedback mechanism to ensure that the
tip and sample only contact intermittently. The imaging force is hence
predominantly vertical, while avoiding strong lateral forces. In
contrast, feedback in contact mode controls the cantilever deflection
rather than the tip height, and this imparts a strong lateral force on
the objects imaged. A schematic of these modes is shown in Figure S3.

\begin{figure}
\includegraphics[width=\columnwidth]{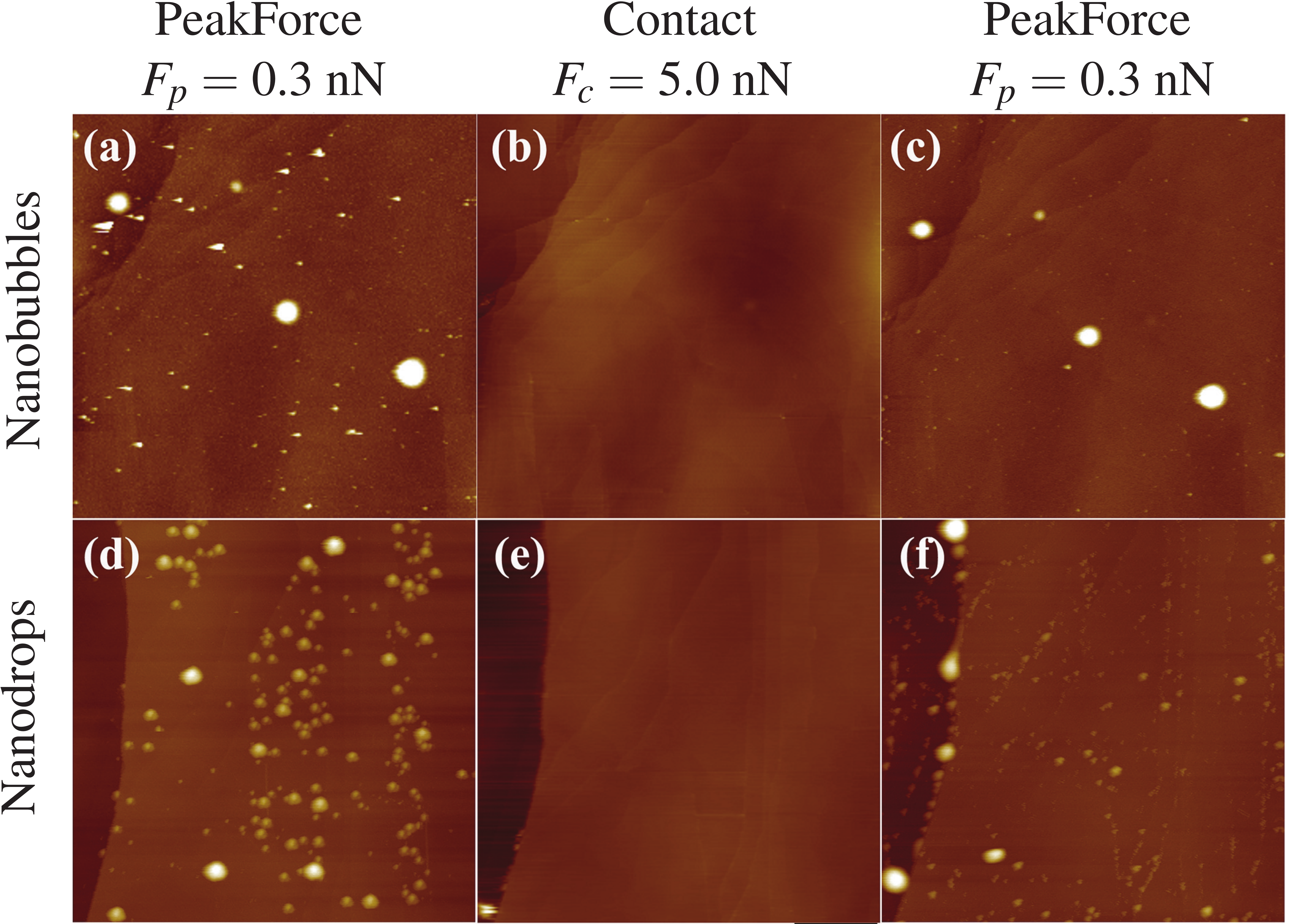}
\caption{\label{fig2} AFM height images of nanobubbles (A-C) and PDMS
  nanodroplets (D-F) before and after an invasive contact mode
  scan. The nanobubbles survive invasive scanning, while the
  nanodroplets' distribution is irreversibly altered, suggesting a
  drastic difference in substrate pinning. (A), (C), (D), and (F) are
  PeakForce mode height images captured at $F_p = $ 0.3 nN. (B) and
  (E) are captured by contact mode imaging (imaging force is $F_c = $
  5 nN). Scan size: 10 $\mu$m $\times$ 5 $\mu$m. Height scale: 50 nm
  for (A-C); 100 nm for (D-F).}
\end{figure}

To observe the influence of strong lateral deformation on nanobubbles
and nanodroplets, we nucleated the objects separately on HOPG
substrate and performed scans in PeakForce mode before (Figures 3A and
D) and after (Figures 3C and F) a contact mode scan with an estimated
imaging force of 5 nN (Figs. 3B and E). Before contact mode imaging,
the deflection sensitivity (in units of m/volts) $S_d$ was
calibrated. The imaging force $F_c$ was estimated as
$F_c = kS_d(a_0 - a_v)$, where $k$ is the spring constant of the
cantilever, $a_0$ the set point and $a_v$ the vertical deflection
signal (both in volts) reading before the AFM was engaged. Figs. 3a-c
show that nanobubbles on the substrate were neither moved nor
destroyed by contact mode scanning. Some irregularly-shaped particles,
likely dirt, were removed by the scanning. We were unable to unpin
nanobubbles from the substrate, even when $F_c$ was increased to
$\approx 50$ nN. On the contrary, the distribution of nanodroplets was
completely altered at $F_c = 5.0$ nN, either by merging with other
nanodroplets or being removed from the scan area (Figure 3F).

To understand why nanodroplets are unpinned and swept away in contact
mode imaging, we make a few estimates. According to Young's equation,
the contact angle of a surface-attached drop is defined only by the
interfacial energies at its three phase line. The tip imparts a steady
lateral force onto the drop, which deforms the drop from one side,
changing its approaching and receding angles (Figure S3). The drop
depins once the difference in angles -- hysteresis -- has reached
a certain amount. Since this threshold is unknown, we instead exploit
the fact that bubbles and drops possess an effective spring constant
\cite{attard_effective_2001} in the same order of magnitude as the
surface tension of the object \cite{attard_direct_2015}. The force
exerted on the drops can then be estimated from the distance of
deformation $x$ to be $F\sim \gamma_{pw}x$, where $\gamma_{pw}$ is the
PDMS-water interface tension. In the experiments, the PDMS-water
interface has a surface tension of 40 mN/m
\cite{ismail_interfacial_2009}, and the nanodroplets have a typical
base radius $\sim$200 nm. Assuming that the maximum deformation of the
drop is in the same order as the base radius, the threshold force to
trigger the unpinning of the drop in contact mode AFM can be estimated
to be 4 nN, in agreement with our experimental observation. On the
other hand, in PeakForce mode, the intermittent contact between tip
and object minimised contact angle hysteresis. Nanodroplets were thus
able to withstand large vertical forces of $\sim$10 nN without
unpinning.

The force threshold required to unpin a nanobubble from its substrate
remains unresolved. If nanobubbles and nanodroplets are pinned to
their substrates by an identical force per unit length, and knowing
that the interfacial energy of water-air is about double that of
water-PDMS (72 versus 40 mJ/m$^2$), the force threshold to unpin a
bubble from the surface in contact mode should be approximately double
that of a PDMS drop with same lateral size, i.e. $F_c \approx 10$
nN. However, we were unable to unpin nanobubbles at any imaging
force in both imaging modes used here. 

\subsection{Force spectroscopy}

\begin{figure}
\includegraphics[width=\columnwidth]{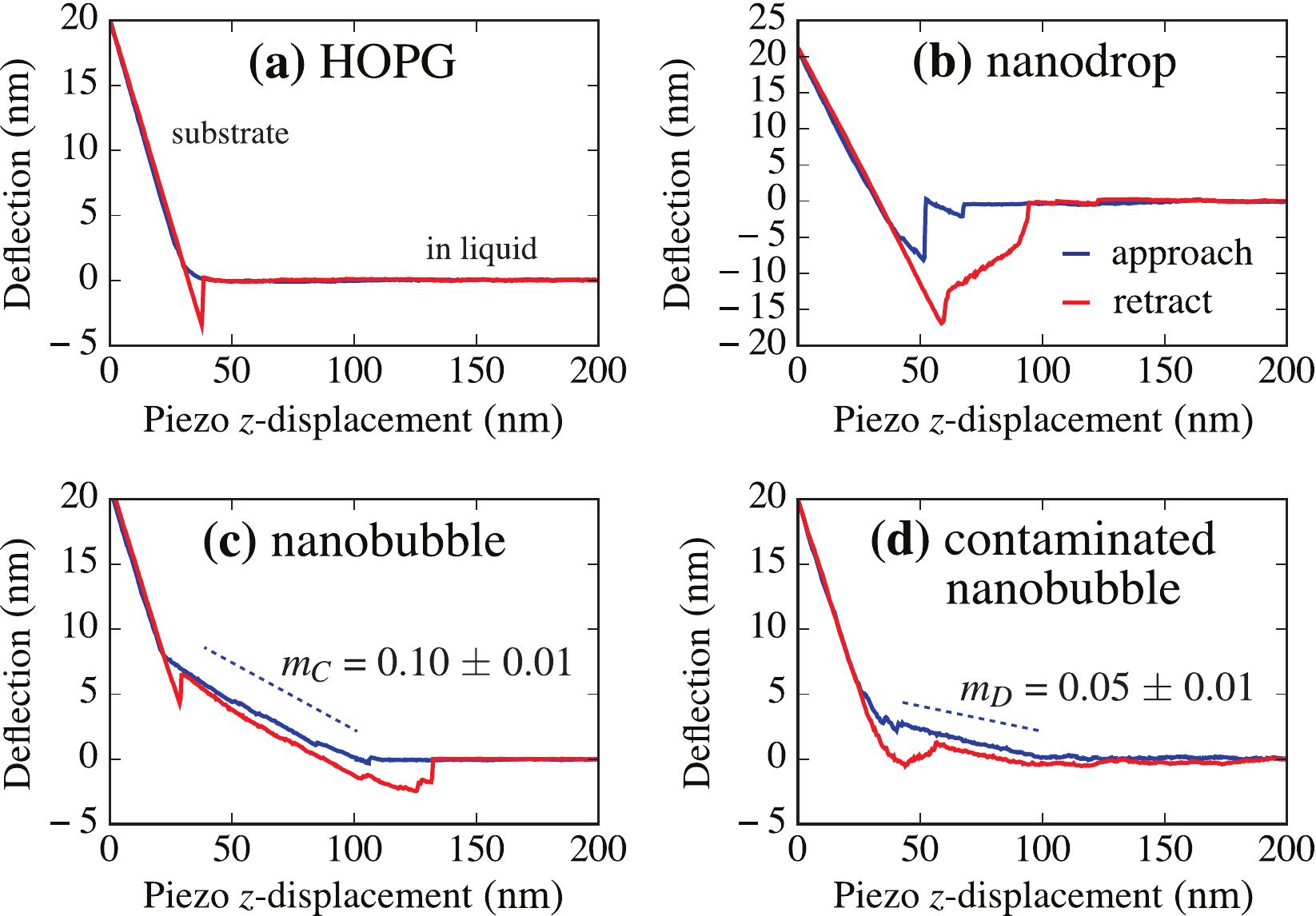}
\caption{\label{fig2} Cantilever deflection versus piezo
  $z$-displacement on the (A) HOPG substrate, (B) PDMS nanodroplet,
  (C) nanobubble and (D) nanobubble contaminated with PDMS. The
  approach curves are given in blue and the retraction curves are
  red. The tip-object slopes $m$ on the approach curves in (C) and (D)
  were calculated with a least-square fit.}
\end{figure}

We also examined the tip-object interaction with AFM force
spectroscopy. The tip-sample interaction is quantified by tracking the
cantilever's vertical deflection as it moves down towards (approach)
and away (retract) from the sample. The force curves are read
right-to-left. Figure 4A shows the deflection-displacement curve on
the bare HOPG substrate. When the tip is far away from the substrate
(`in liquid'), the deflection is zero. The deflection then increases
linearly as the tip comes into contact with the HOPG substrate
(`substrate'); the slope of the deflection-displacement curve gives a
measure of the effective stiffness of the substrate.

We next show the force curves when the tip comes into contact first
with a nanodroplet (Figure 4B) or a nanobubble (Figure 4C) before
contacting the substrate.The tip-bubble interaction is characterised
by a linear slope in the deflection-piezo displacement curve, which is
consistent with theory \cite{attard_effective_2001} and experiments
\cite{zhang_physical_2006, an_surface_2015}. In our limited testing,
this linear dependence occurs independent of the type of salt used in
the electrolysis and probe wettability. Theory suggests that the
tip-bubble interaction is linear \cite{attard_direct_2015} provided
the magnitude of deformation by the tip is within a compliance regime
\cite{attard_effective_2001, chan2001forces}. We assume that
nanobubbles fall within this regime because the maximum deformation of
a spherically-capped nanobubble is its height $h \sim 50$ nm, which is
small compared to its radius of curvature $R_c \sim 1$ $\mu$m. We note
the caveat that in very pure systems, the initial contact between the
AFM tip and the bubble is marked with a pronounced snap-in
\cite{zhang_physical_2006}, but this is suppressed in our case,
perhaps due to disruption of electric double layer forces by
electrolyte ions. However, the snap-in itself does not affect the
linear dependence of the bubble-tip interaction. On the other hand,
the tip-nanodroplet interaction is non-linear. Sharp kinks are
observed in the approach curve in Figure 4b, representing a
jump-to-contact with the substrate when the gradient of van der Waals
forces between the oil and the tip overcomes the cantilever's spring
constant \cite{connell_investigating_2002}.

We were also able to distinguish between pure nanobubble and one that
that was contaminated with a layer of oil. As with a pure nanobubble,
the tip-contaminated bubble interaction is linear, but with a slope 
much smaller than for the pure nanobubble. Theory suggests that
the slope of the deflection-displacement curve $m$ is proportional to
the surface tension of the interface $\gamma$
\cite{attard_direct_2015}. This agrees well with our experimental
observation that the ratio of the deflection curves on the
contaminated bubble compared to the pure bubble (Figures 3C-D),
$0.05/0.10$, is approximately the ratio of the surface tensions of
water and PDMS,
$m_C/m_D \approx \gamma_{\rm PDMS}/\gamma_{\rm water} = 5/9$.

\section{Conclusions}
In this Article, we identify a number of robust indicators to
distinguish between nanobubbles and nanodroplets when they are imaged
in AFM, such as differences in contact angles, force curves and
response to different types of imaging forces. Our central result is
that bubbles and drops exhibit very different responses under vertical
and lateral forces. In PeakForce mode, we find that nanobubbles
disappear cleanly from the AFM height images under large vertical
imaging forces, but nanodroplets maintain a nanometric molecular layer
under imaging forces in the order of tens of nN. In the standard
contact mode AFM, we observe that nanobubbles are very strongly pinned
onto their substrates and survive violent, lateral imaging forces of
up to 50 nN without being moved or destroyed. Nanodroplets, on the
other hand, are weakly pinned to their substrates and are easily
depinned during contact mode imaging.

Our findings address a very urgent need in the field of nanobubbles to
distinguish between gaseous objects and polymeric contamination which
occurs easily during nucleation experiments. Although other methods of
distinguishing between nanobubbles and contamination already exist,
these tests are destructive and require specialised optical techniques
inaccessible to the majority of research groups which image
nanobubbles with AFM. We anticipate that our Article will equip
researchers and engineers in the field of nanobubbles with a new
perspective to judge previous work, as well as a diagnostic to verify
the quality of their existing nucleation protocols.

\section{Materials and methods}

\subsection{Nucleation of nanobubbles and nanodroplets}

Hydrogen nanobubbles were generated by the electrolysis of aqueous
solutions on highly ordered pyrolytic graphite (HOPG) substrate, the
details of which are well-documented in previous work
\cite{wu2008cleaning, yang2009electrolytically}. Here, HOPG was used
as the cathode and a copper wire as an anode within sodium phosphate
solution (10 mM), and a DC voltage of 1.5 V was applied for 90 s. 

To create oil nanodroplets, polydimethylsiloxane (PDMS) (Sylgard 184,
Dow Corning, United States) was diluted with de-ionized water to
1:10,000 (vol/vol) (as described in \cite{berkelaar_exposing_2014})
and then directly deposited on HOPG.

\subsection{AFM imaging}

Atomic force microscopy scans were captured with a Bioscope Catalyst AFM (Bruker
Corporation, United States). A V-shaped cantilever (SNL-10 A, Bruker)
with a nominal spring constant of $k =$ 0.35 N m$^{-1}$ was used in an
open liquid system. Before experiments, the AFM tip was treated with
oxygen plasma for 15 s to render it hydrophilic, and the cantilever's
spring constant was calibrated by a thermal method using built-in
software. 

To provide a baseline distribution of nanobubbles and nanodroplets
that can be compared with previous work (Figure 1), nanobubbles and
nanodroplets were independently imaged in the commonly-used tapping
mode at a setpoint ratio of $A/A_0 = 0.8$, where $A_0 \approx 10$ nm
is the free amplitude of the cantilever.

We studied the influence of primarily vertical imaging forces on the
objects (Figure 2) by imaging in PeakForce tapping mode. In PeakForce
mode, the cantilever is oscillated at a machine-selected frequency of
about 1-2 kHz, far below its resonance in water ($\sim 30-40$ kHz)
\cite{kaemmer2011}. 

\subsection{Calculation of contact angle}
All contact angles reported in the manuscript are calculated with a
custom algorithm in Python with the scipy package. The left and right
boundaries of the bubble were determined using a peak finding
algorithm \cite{Duarte2015}, before all the height points between the
two detected boundaries were fitted using a standard least squares
algorithm to a circle (spherical cap in axis symmetry). Finally, to
determine the contact angle, a chord was constructed by interpolating
points on the substrate outside the bubble. From the position of this
chord relative to the circle, an elementary geometrical relation
yields the contact angle.

\begin{acknowledgement}
  We acknowledge funding from a competitive research programme under
  the auspices of the Singapore government's National Research
  Foundation (programme no. NRF-CRP9-2011-04). B.H.T. acknowledges
  financial support from the Agency of Science, Technology and
  Research in Singapore. 
\end{acknowledgement}

\begin{suppinfo}

The Supporting Information consists of a single file with four
additional figures and one table. 

% A listing of the contents of each file supplied as Supporting Information
% should be included. For instructions on what should be included in the
% Supporting Information as well as how to prepare this material for
% publications, refer to the journal's Instructions for Authors.

% The following files are available free of charge.
% \begin{itemize}
%   \item Filename: brief description
%   \item Filename: brief description
% \end{itemize}

\end{suppinfo}

\bibliography{ref}

\providecommand{\noopsort}[1]{}\providecommand{\singleletter}[1]{#1}%
\providecommand{\latin}[1]{#1}
\providecommand*\mcitethebibliography{\thebibliography}
\csname @ifundefined\endcsname{endmcitethebibliography}
  {\let\endmcitethebibliography\endthebibliography}{}
\begin{mcitethebibliography}{32}
\providecommand*\natexlab[1]{#1}
\providecommand*\mciteSetBstSublistMode[1]{}
\providecommand*\mciteSetBstMaxWidthForm[2]{}
\providecommand*\mciteBstWouldAddEndPuncttrue
  {\def\EndOfBibitem{\unskip.}}
\providecommand*\mciteBstWouldAddEndPunctfalse
  {\let\EndOfBibitem\relax}
\providecommand*\mciteSetBstMidEndSepPunct[3]{}
\providecommand*\mciteSetBstSublistLabelBeginEnd[3]{}
\providecommand*\EndOfBibitem{}
\mciteSetBstSublistMode{f}
\mciteSetBstMaxWidthForm{subitem}{(\alph{mcitesubitemcount})}
\mciteSetBstSublistLabelBeginEnd
  {\mcitemaxwidthsubitemform\space}
  {\relax}
  {\relax}

\bibitem[Craig(2010)]{craig_very_2010}
Craig,~V. S.~J. Very small bubbles at surfaces--the nanobubble puzzle.
  \emph{Soft Matter} \textbf{2010}, \emph{7}, 40--48\relax
\mciteBstWouldAddEndPuncttrue
\mciteSetBstMidEndSepPunct{\mcitedefaultmidpunct}
{\mcitedefaultendpunct}{\mcitedefaultseppunct}\relax
\EndOfBibitem
\bibitem[Seddon and Lohse(2011)Seddon, and Lohse]{seddon_nanobubbles_2011}
Seddon,~J. R.~T.; Lohse,~D. Nanobubbles and micropancakes: gaseous domains on
  immersed substrates. \emph{J. Phys.: Condens. Matter} \textbf{2011},
  \emph{23}, 133001\relax
\mciteBstWouldAddEndPuncttrue
\mciteSetBstMidEndSepPunct{\mcitedefaultmidpunct}
{\mcitedefaultendpunct}{\mcitedefaultseppunct}\relax
\EndOfBibitem
\bibitem[Lohse and Zhang(2015)Lohse, and Zhang]{lohse_surface_2015}
Lohse,~D.; Zhang,~X. Surface nanobubbles and nanodroplets. \emph{Rev. Mod.
  Phys.} \textbf{2015}, \emph{87}, 981--1035\relax
\mciteBstWouldAddEndPuncttrue
\mciteSetBstMidEndSepPunct{\mcitedefaultmidpunct}
{\mcitedefaultendpunct}{\mcitedefaultseppunct}\relax
\EndOfBibitem
\bibitem[Zhang \latin{et~al.}(2015)Zhang, Lu, Tan, Bao, He, Sun, and
  Lohse]{zhang_formation_2015}
Zhang,~X.; Lu,~Z.; Tan,~H.; Bao,~L.; He,~Y.; Sun,~C.; Lohse,~D. Formation of
  surface nanodroplets under controlled flow conditions. \emph{Proc. Natl.
  Acad. Sci. U.S.A.} \textbf{2015}, \emph{112}, 9253--9257\relax
\mciteBstWouldAddEndPuncttrue
\mciteSetBstMidEndSepPunct{\mcitedefaultmidpunct}
{\mcitedefaultendpunct}{\mcitedefaultseppunct}\relax
\EndOfBibitem
\bibitem[Karatay \latin{et~al.}(2013)Karatay, Haase, Visser, Sun, Lohse, Tsai,
  and Lammertink]{karatay_control_2013}
Karatay,~E.; Haase,~A.~S.; Visser,~C.~W.; Sun,~C.; Lohse,~D.; Tsai,~P.~A.;
  Lammertink,~R. G.~H. Control of slippage with tunable bubble mattresses.
  \emph{Proc. Natl. Acad. Sci. U.S.A.} \textbf{2013}, \emph{110},
  8422--8426\relax
\mciteBstWouldAddEndPuncttrue
\mciteSetBstMidEndSepPunct{\mcitedefaultmidpunct}
{\mcitedefaultendpunct}{\mcitedefaultseppunct}\relax
\EndOfBibitem
\bibitem[Zhang and Ducker(2007)Zhang, and Ducker]{zhang_formation_2007}
Zhang,~X.~H.; Ducker,~W. Formation of {Interfacial} {Nanodroplets} through
  {Changes} in {Solvent} {Quality}. \emph{Langmuir} \textbf{2007}, \emph{23},
  12478--12480\relax
\mciteBstWouldAddEndPuncttrue
\mciteSetBstMidEndSepPunct{\mcitedefaultmidpunct}
{\mcitedefaultendpunct}{\mcitedefaultseppunct}\relax
\EndOfBibitem
\bibitem[German \latin{et~al.}(2014)German, Wu, An, Craig, Mega, and
  Zhang]{german_interfacial_2014}
German,~S.~R.; Wu,~X.; An,~H.; Craig,~V. S.~J.; Mega,~T.~L.; Zhang,~X.
  Interfacial {Nanobubbles} {Are} {Leaky}: {Permeability} of the {Gas}/{Water}
  {Interface}. \emph{ACS Nano} \textbf{2014}, \emph{8}, 6193--6201\relax
\mciteBstWouldAddEndPuncttrue
\mciteSetBstMidEndSepPunct{\mcitedefaultmidpunct}
{\mcitedefaultendpunct}{\mcitedefaultseppunct}\relax
\EndOfBibitem
\bibitem[Chan and Ohl(2012)Chan, and Ohl]{chan_tirf_2012}
Chan,~C.~U.; Ohl,~C.-D. Total-{Internal}-{Reflection}-{Fluorescence}
  {Microscopy} for the {Study} of {Nanobubble} {Dynamics}. \emph{Phys. Rev.
  Lett.} \textbf{2012}, \emph{109}, 174501\relax
\mciteBstWouldAddEndPuncttrue
\mciteSetBstMidEndSepPunct{\mcitedefaultmidpunct}
{\mcitedefaultendpunct}{\mcitedefaultseppunct}\relax
\EndOfBibitem
\bibitem[Berkelaar \latin{et~al.}(2014)Berkelaar, Dietrich, Kip, Kooij,
  Zandvliet, and Lohse]{berkelaar_exposing_2014}
Berkelaar,~R.~P.; Dietrich,~E.; Kip,~G. A.~M.; Kooij,~E.~S.; Zandvliet,~H.
  J.~W.; Lohse,~D. Exposing nanobubble-like objects to a degassed environment.
  \emph{Soft Matter} \textbf{2014}, \emph{10}, 4947--4955\relax
\mciteBstWouldAddEndPuncttrue
\mciteSetBstMidEndSepPunct{\mcitedefaultmidpunct}
{\mcitedefaultendpunct}{\mcitedefaultseppunct}\relax
\EndOfBibitem
\bibitem[An \latin{et~al.}(2014)An, Liu, and Craig]{an_wetting_2014}
An,~H.; Liu,~G.; Craig,~V. S.~J. Wetting of nanophases: {Nanobubbles},
  nanodroplets and micropancakes on hydrophobic surfaces. \emph{Adv. Colloid
  Interfac.} \textbf{2014}, \emph{222}, 9--12\relax
\mciteBstWouldAddEndPuncttrue
\mciteSetBstMidEndSepPunct{\mcitedefaultmidpunct}
{\mcitedefaultendpunct}{\mcitedefaultseppunct}\relax
\EndOfBibitem
\bibitem[Carr \latin{et~al.}(2012)Carr, Nalwa, Mahadevapuram, Chen, Anderegg,
  and Chaudhary]{carr_plastic-syringe_2012}
Carr,~J.~A.; Nalwa,~K.~S.; Mahadevapuram,~R.; Chen,~Y.; Anderegg,~J.;
  Chaudhary,~S. Plastic-{Syringe} {Induced} {Silicone} {Contamination} in
  {Organic} {Photovoltaic} {Fabrication}: {Implications} for {Small}-{Volume}
  {Additives}. \emph{ACS Appl. Mater. Interfaces} \textbf{2012}, \emph{4},
  2831--2835\relax
\mciteBstWouldAddEndPuncttrue
\mciteSetBstMidEndSepPunct{\mcitedefaultmidpunct}
{\mcitedefaultendpunct}{\mcitedefaultseppunct}\relax
\EndOfBibitem
\bibitem[Buettner \latin{et~al.}(1991)Buettner, Scott, Kerber, and
  M{\"u}gge]{buettner1991free}
Buettner,~G.~R.; Scott,~B.~D.; Kerber,~R.~E.; M{\"u}gge,~A. Free radicals from
  plastic syringes. \emph{Free Rad. Bio. Med.} \textbf{1991}, \emph{11},
  69--70\relax
\mciteBstWouldAddEndPuncttrue
\mciteSetBstMidEndSepPunct{\mcitedefaultmidpunct}
{\mcitedefaultendpunct}{\mcitedefaultseppunct}\relax
\EndOfBibitem
\bibitem[Siniawski \latin{et~al.}(2015)Siniawski, Felts, Kurilich, Lopez, and
  Malik]{siniawski_method_2015}
Siniawski,~M.~T.; Felts,~J.; Kurilich,~D.; Lopez,~A.; Malik,~A. Method for
  testing sliding frictional response of lubricious thin films used in plastic
  medical syringes. \emph{Tribology S} \textbf{2015}, \relax
\mciteBstWouldAddEndPunctfalse
\mciteSetBstMidEndSepPunct{\mcitedefaultmidpunct}
{}{\mcitedefaultseppunct}\relax
\EndOfBibitem
\bibitem[Curtis and Colas(2013)Curtis, and Colas]{curtis_chapter_2013}
Curtis,~J.; Colas,~A. In \emph{Biomaterials {Science} ({Third} {Edition})};
  Ratner,~B.~D., Ed.; Academic Press, 2013; pp 1106--1116\relax
\mciteBstWouldAddEndPuncttrue
\mciteSetBstMidEndSepPunct{\mcitedefaultmidpunct}
{\mcitedefaultendpunct}{\mcitedefaultseppunct}\relax
\EndOfBibitem
\bibitem[Chan \latin{et~al.}(2015)Chan, Chen, Arora, and
  Ohl]{chan_collapse_2015}
Chan,~C.~U.; Chen,~L.; Arora,~M.; Ohl,~C.-D. Collapse of {Surface}
  {Nanobubbles}. \emph{Phys. Rev. Lett.} \textbf{2015}, \emph{114},
  114505\relax
\mciteBstWouldAddEndPuncttrue
\mciteSetBstMidEndSepPunct{\mcitedefaultmidpunct}
{\mcitedefaultendpunct}{\mcitedefaultseppunct}\relax
\EndOfBibitem
\bibitem[Seo \latin{et~al.}(2015)Seo, German, Mega, and Ducker]{seo_phase_2015}
Seo,~D.; German,~S.~R.; Mega,~T.~L.; Ducker,~W.~A. The {Phase} {State} of
  {Interfacial} {Nanobubbles}. \emph{J. Phys. Chem. C} \textbf{2015},
  \emph{119}, 14262--14266\relax
\mciteBstWouldAddEndPuncttrue
\mciteSetBstMidEndSepPunct{\mcitedefaultmidpunct}
{\mcitedefaultendpunct}{\mcitedefaultseppunct}\relax
\EndOfBibitem
\bibitem[Yang \latin{et~al.}(2013)Yang, Lu, and Hwang]{yang_imaging_2013}
Yang,~C.-W.; Lu,~Y.-H.; Hwang,~I.-S. Imaging surface nanobubbles at
  graphite-water interfaces with different atomic force microscopy modes.
  \emph{J. Phys.: Condens. Matter} \textbf{2013}, \emph{25}, 184010\relax
\mciteBstWouldAddEndPuncttrue
\mciteSetBstMidEndSepPunct{\mcitedefaultmidpunct}
{\mcitedefaultendpunct}{\mcitedefaultseppunct}\relax
\EndOfBibitem
\bibitem[Walczyk \latin{et~al.}(2013)Walczyk, Sch{\"o}n, and
  Sch{\"o}nherr]{walczyk2013effect}
Walczyk,~W.; Sch{\"o}n,~P.~M.; Sch{\"o}nherr,~H. The effect of PeakForce
  tapping mode AFM imaging on the apparent shape of surface nanobubbles.
  \emph{J. Phys. Cond. Mat.} \textbf{2013}, \emph{25}, 184005\relax
\mciteBstWouldAddEndPuncttrue
\mciteSetBstMidEndSepPunct{\mcitedefaultmidpunct}
{\mcitedefaultendpunct}{\mcitedefaultseppunct}\relax
\EndOfBibitem
\bibitem[Zhao \latin{et~al.}(2013)Zhao, Song, Wang, Dai, Zhang, Dong, L\"{u},
  and Hu]{zhao_mechanical_2013}
Zhao,~B.; Song,~Y.; Wang,~S.; Dai,~B.; Zhang,~L.; Dong,~Y.; L\"{u},~J.; Hu,~J.
  Mechanical mapping of nanobubbles by {PeakForce} atomic force microscopy.
  \emph{Soft Matter} \textbf{2013}, \emph{9}, 8837--8843\relax
\mciteBstWouldAddEndPuncttrue
\mciteSetBstMidEndSepPunct{\mcitedefaultmidpunct}
{\mcitedefaultendpunct}{\mcitedefaultseppunct}\relax
\EndOfBibitem
\bibitem[Heslot \latin{et~al.}(1989)Heslot, Fraysse, and
  Cazabat]{heslot_molecular_1989}
Heslot,~F.; Fraysse,~N.; Cazabat,~A.~M. Molecular layering in the spreading of
  wetting liquid drops. \emph{Nature} \textbf{1989}, \emph{338}, 640--642\relax
\mciteBstWouldAddEndPuncttrue
\mciteSetBstMidEndSepPunct{\mcitedefaultmidpunct}
{\mcitedefaultendpunct}{\mcitedefaultseppunct}\relax
\EndOfBibitem
\bibitem[Kaemmer(2011)]{kaemmer2011}
Kaemmer,~S.~B. Application note 133: introduction to Bruker's ScanAsyst and
  PeakForce Tapping AFM Technology. Bruker Corporation, 2011\relax
\mciteBstWouldAddEndPuncttrue
\mciteSetBstMidEndSepPunct{\mcitedefaultmidpunct}
{\mcitedefaultendpunct}{\mcitedefaultseppunct}\relax
\EndOfBibitem
\bibitem[Attard and Miklavcic(2001)Attard, and
  Miklavcic]{attard_effective_2001}
Attard,~P.; Miklavcic,~S.~J. Effective {Spring} {Constant} of {Bubbles} and
  {Droplets}. \emph{Langmuir} \textbf{2001}, \emph{17}, 8217--8223\relax
\mciteBstWouldAddEndPuncttrue
\mciteSetBstMidEndSepPunct{\mcitedefaultmidpunct}
{\mcitedefaultendpunct}{\mcitedefaultseppunct}\relax
\EndOfBibitem
\bibitem[Attard(2015)]{attard_direct_2015}
Attard,~P. Direct {Measurement} of the {Surface} {Tension} of {Nanobubbles}.
  \emph{arXiv:1505.02217 [cond-mat, physics]} \textbf{2015}, \relax
\mciteBstWouldAddEndPunctfalse
\mciteSetBstMidEndSepPunct{\mcitedefaultmidpunct}
{}{\mcitedefaultseppunct}\relax
\EndOfBibitem
\bibitem[Ismail \latin{et~al.}(2009)Ismail, Grest, Heine, Stevens, and
  Tsige]{ismail_interfacial_2009}
Ismail,~A.~E.; Grest,~G.~S.; Heine,~D.~R.; Stevens,~M.~J.; Tsige,~M.
  Interfacial {Structure} and {Dynamics} of {Siloxane} {Systems}:
  {PDMS}-{Vapor} and {PDMS}-{Water}. \emph{Macromolecules} \textbf{2009},
  \emph{42}, 3186--3194\relax
\mciteBstWouldAddEndPuncttrue
\mciteSetBstMidEndSepPunct{\mcitedefaultmidpunct}
{\mcitedefaultendpunct}{\mcitedefaultseppunct}\relax
\EndOfBibitem
\bibitem[Zhang \latin{et~al.}(2006)Zhang, Maeda, and
  Craig]{zhang_physical_2006}
Zhang,~X.~H.; Maeda,~N.; Craig,~V. S.~J. Physical {Properties} of {Nanobubbles}
  on {Hydrophobic} {Surfaces} in {Water} and {Aqueous} {Solutions}.
  \emph{Langmuir} \textbf{2006}, \emph{22}, 5025--5035\relax
\mciteBstWouldAddEndPuncttrue
\mciteSetBstMidEndSepPunct{\mcitedefaultmidpunct}
{\mcitedefaultendpunct}{\mcitedefaultseppunct}\relax
\EndOfBibitem
\bibitem[An \latin{et~al.}(2015)An, Liu, Atkin, and Craig]{an_surface_2015}
An,~H.; Liu,~G.; Atkin,~R.; Craig,~V. S.~J. Surface {Nanobubbles} in
  {Nonaqueous} {Media}: {Looking} for {Nanobubbles} in {DMSO}, {Formamide},
  {Propylene} {Carbonate}, {Ethylammonium} {Nitrate}, and {Propylammonium}
  {Nitrate}. \emph{ACS Nano} \textbf{2015}, \emph{9}, 7596--7607\relax
\mciteBstWouldAddEndPuncttrue
\mciteSetBstMidEndSepPunct{\mcitedefaultmidpunct}
{\mcitedefaultendpunct}{\mcitedefaultseppunct}\relax
\EndOfBibitem
\bibitem[Chan \latin{et~al.}(2001)Chan, Dagastine, and White]{chan2001forces}
Chan,~D.; Dagastine,~R.; White,~L. Forces between a rigid probe particle and a
  liquid interface: I. The repulsive case. \emph{J. Colloid Interfac. Sci.}
  \textbf{2001}, \emph{236}, 141--154\relax
\mciteBstWouldAddEndPuncttrue
\mciteSetBstMidEndSepPunct{\mcitedefaultmidpunct}
{\mcitedefaultendpunct}{\mcitedefaultseppunct}\relax
\EndOfBibitem
\bibitem[Connell \latin{et~al.}(2002)Connell, Allen, Roberts, Davies, Davies,
  Tendler, and Williams]{connell_investigating_2002}
Connell,~S. D.~A.; Allen,~S.; Roberts,~C.~J.; Davies,~J.; Davies,~M.~C.;
  Tendler,~S. J.~B.; Williams,~P.~M. Investigating the {Interfacial}
  {Properties} of {Single}-{Liquid} {Nanodroplets} by {Atomic} {Force}
  {Microscopy}. \emph{Langmuir} \textbf{2002}, \emph{18}, 1719--1728\relax
\mciteBstWouldAddEndPuncttrue
\mciteSetBstMidEndSepPunct{\mcitedefaultmidpunct}
{\mcitedefaultendpunct}{\mcitedefaultseppunct}\relax
\EndOfBibitem
\bibitem[Wu \latin{et~al.}(2008)Wu, Chen, Dong, Mao, Sun, Chen, Craig, and
  Hu]{wu2008cleaning}
Wu,~Z.; Chen,~H.; Dong,~Y.; Mao,~H.; Sun,~J.; Chen,~S.; Craig,~V.~S.; Hu,~J.
  Cleaning using nanobubbles: defouling by electrochemical generation of
  bubbles. \emph{J. Colloid Interfac. Sci.} \textbf{2008}, \emph{328},
  10--14\relax
\mciteBstWouldAddEndPuncttrue
\mciteSetBstMidEndSepPunct{\mcitedefaultmidpunct}
{\mcitedefaultendpunct}{\mcitedefaultseppunct}\relax
\EndOfBibitem
\bibitem[Yang \latin{et~al.}(2009)Yang, Tsai, Kooij, Prosperetti, Zandvliet,
  and Lohse]{yang2009electrolytically}
Yang,~S.; Tsai,~P.; Kooij,~E.~S.; Prosperetti,~A.; Zandvliet,~H.~J.; Lohse,~D.
  Electrolytically generated nanobubbles on highly orientated pyrolytic
  graphite surfaces. \emph{Langmuir} \textbf{2009}, \emph{25}, 1466--1474\relax
\mciteBstWouldAddEndPuncttrue
\mciteSetBstMidEndSepPunct{\mcitedefaultmidpunct}
{\mcitedefaultendpunct}{\mcitedefaultseppunct}\relax
\EndOfBibitem
\bibitem[Duarte(2015)]{Duarte2015}
Duarte,~M. Notes on Scientific Computing for Biomechanics and Motor Control.
  \url{https://github.com/demotu/BMC}, 2015\relax
\mciteBstWouldAddEndPuncttrue
\mciteSetBstMidEndSepPunct{\mcitedefaultmidpunct}
{\mcitedefaultendpunct}{\mcitedefaultseppunct}\relax
\EndOfBibitem
\end{mcitethebibliography}

\end{document}